\documentclass[apj]{emulateapj}
\usepackage{multirow}
\usepackage{booktabs}
\usepackage{graphicx}
\usepackage{amsmath}
\usepackage{longtable}
\usepackage{rotating}
\usepackage{enumerate}
\usepackage{color}

\newcommand{\mth}{M_{\rm th}}

\newcommand{\mrm}[1]{\mathrm{#1}}

\begin{document}
\title{The Observability of Vortex-Driven Spiral Arms in Protoplanetary Disk: Basic Spiral Properties}

\author{Pinghui Huang}
\affiliation{CAS Key Laboratory of Planetary Sciences, Purple Mountain Observatory, Chinese Academy of Sciences, Nanjing 210008, China}
\affiliation{University of Chinese Academy of Sciences, Beijing 100049, China}
\affiliation{Theoretical Division, Los Alamos National Laboratory, Los Alamos, NM 87545, USA}
\author{Ruobing Dong}
\affiliation{Department of Physics \& Astronomy, University of Victoria, Victoria, BC, Canada}
\author{Hui Li, Shengtai Li}
\affiliation{Theoretical Division, Los Alamos National Laboratory, Los Alamos, NM 87545, USA}
\author{Jianghui Ji}
\affiliation{CAS Key Laboratory of Planetary Sciences, Purple Mountain Observatory, Chinese Academy of Sciences, Nanjing 210008, China}
\affiliation{CAS Center for Excellence in Comparative Planetology, Hefei 230026, China}

\altaffiltext{1}{}

\begin{abstract}
Some circumstellar disks are observed to show prominent spiral arms in infrared scattered light or (sub-)millimeter dust continuum. The spirals might be formed from self-gravity, shadows, or planet-disk interactions. Recently, it was hypothesized that massive vortices can drive spiral arms in protoplanetary disks in a way analogous to planets. In this paper, we study the basic properties of vortex-driven spirals by the Rossby Wave Instability in 2D hydrodynamics simulations. We study how the surface density contrast, the number, and the shape of vortex-driven spirals depend on the properties of the vortex. We also compare vortex-driven spirals with those induced by planets. The surface density contrast of vortex-driven spirals in our simulations are comparable to those driven by a sub-thermal mass planet, typically a few to a few tens of Earth masses. In addition, different from the latter, the former is not sensitive to the mass of the vortex. Vortex-driven spiral arms are not expected to be detectable in current scattered light observations, and the prominent spirals observed in scattered light in a few protoplanetary disks, such as SAO~206462 (HD~135344B), MWC~758, and LkH$\alpha$~330, are unlikely to be induced by the candidate vortices in them.
\end{abstract}



\section{Introduction}\label{sec:intro}
In recent years, many circumstellar disks were resolved by high angular resolution infrared and (sub-)millimeter observations~\citep{brogan20152014,long2018gaps,andrews2018disk,avenhaus2018dartts,liu2016circumstellar,huang2018disk2,huang2018disk3}. These disks present a large diversity in morphology, showing rings, cavities, spiral arms, and dust crescents, likely produced by planets~\citep{ou2007disk,fung2014empty,zhu2014dust,jin2016modeling,liu2018new,van2018rings,jin2019new}. Specifically, several disks present both dust crescents in continuum emission and spirals in scattered light, such as MWC~758~\citep{isella2010millimeter, boehler2018complex, dong2018eccentric,grady2012spiral,benisty2015asymmetric}, SAO~206462 (HD~135344B)~\citep{van2016vortices,perez2014large,muto2012discovery,garufi2013small}, LkH$\alpha$~330~\citep{isella2013azimuthal,akiyama2016spiral,uyama2018subaru}, V1247~Ori~\citep{ohta2016extreme,kraus2017dust}, HD~142527~\citep{casassus2013flows,canovas2013near,avenhaus2014structures}, and AB~Aur~\citep{hashimoto2011direct,tang2017planet}.

Crescents in continuum emission have been proposed to be dust trapping in vortices generated by Rossby Wave Instability~\cite[][RWI]{lovelace1999rossby,li2000rossby,li2001rossby}. The RWI can be triggered by local Rossby wave trapped in a steep density bump, either at the edges of planet-opened gaps or at a dead zone edge~\citep{lin2014testing,miranda2017long}. Spiral features have been proposed to be the density waves excited by planets~\citep{dong2015observational} or by the gravitational instability~\citep{kratter2016gravitational,dong2015spiral}. They may also be produced by shadows~\citep{benisty2017shadows,benisty2018shadows,montesinos2018planetary}. \cite{van2016vortices} and \cite{cazzoletti2018evidence} highlighted the connection between the two-arm spirals and the dust crescent seen in the SAO~206462 disk. They proposed that the dust crescent is a massive vortex, and it is exciting the observed spirals in a way similar to a planet with a similar mass.

In this paper, we study the properties of spirals induced by a large vortex generated by the RWI. \S\ref{sec:setup} shows the numerical setup of our hydrodynamic models. \S\ref{sec:hydro_results} and \S\ref{sec:properties} present hydrodynamic results and the properties of vortices induced spiral arms. We will summarize our conclusions in \S\ref{sec:summary}.

\section{Simulations}\label{sec:setup}
We carry out 2D global hydrodynamical simulations in polar coordinate using the LA-COMPASS code (Los Alamos COMPutional Astrophysics Simulation Suite)~\citep{li2005potential,li2008type,fu2014effects} to follow the evolution of gas in a protoplanetary disk. The code units of mass, length, and time are the star mass $M_\star = 1\;\mrm{M_\sun}$, $R_0 = 50\;\mrm{au}$, and the dynamical timescale at $R_0$, i.e. $\tau_\mrm{dyn,0} = \Omega_{K,0}^{-1} = \left(R_0^3/GM_\star\right)^{1/2}$, respectively. The equation of the state is locally isothermal $P_\mrm{g} = c_\mrm{s}^2 \Sigma_\mrm{g}$, where $P_\mrm{g}$ is the vertically integrated pressure, and $c_\mrm{s}$ and $\Sigma_\mrm{g}$ are the sound speed and the surface density of gas, respectively. The behaviours of density waves in linear regime are insensitive to the EOS~\citep{dong2011density1,miranda2019gaps}, while the spiral shock location and wave amplitude depend on the EOS in non-linear regime ~\citep{goodman2001planetary,dong2011density2}. The distribution of $c_\mrm{s}$ and $\Sigma_\mrm{g}$ are $c_\mrm{s}\left(R\right) = c_\mrm{s,0}\left(R/R_0\right)^{-1/4}$ and $\Sigma_\mrm{g} = \Sigma_0\left(R/R_0\right)^{-1}$, with $\Sigma_0$ normalized by the total disk mass $M_{\rm disk}$. The scale height profile is chosen as $H/R = 0.05\left(R/R_0\right)^{1/4}$.

The inner and outer boundaries of our simulations are $R_\mrm{min} = 0.2R_0 = 10\;\mrm{au}$ and $R_\mrm{max} = 4.5R_0 = 225\;\mrm{au}$. We used the outflow and inflow boundary conditions for the inner and outer boundaries, respectively. The simulations are carried on a linear-grid. The resolutions are $2048\times3072$ along the $R$ and $\Phi$ (azimuthal) directions. The scale height at $R_0$ is resolved by 25 cells along the radial direction.

To avoid the disturbance of planet-induced density waves, we choose the dead zone edge model (no planet in simulations) to generate vortices and spirals~\citep{miranda2017long}. The dead zone is a region with low viscosity and consequently weak accretion. Viscosity in PPDs come from turbulence, which are likely produced by instabilities, such as the Magnetorotational Instability~\citep{balbus1998instability}. A sufficiently ionized disk threaded by magnetic fields in the ideal MHD regime can be MRI active. However, at tens of AU the midplane of PPDs are subject to non-ideal MHD effects such as the Ohmic resistivity, Hall effect, and in particular Ambipolar diffusion. As a result the MRI is not active, and PPDs have a ``dead zone'' at the midplane with weak turbulence, thus low viscosity~\citep{gammie1996layered, bai2011ambipolar,bai2014hall1,bai2015hall2}. The outer disk beyond the dead zone may be MRI active due to cosmic ray ionization~\citep{armitage2011dynamics}. Here we adopt the toy model of the dead zone edge as described by ~\cite{regaly2011possible}:
\begin{equation}
    \alpha\left(R\right) = \alpha_0 - \frac{\alpha_0 - \alpha_\mrm{DZ}}{2}\left[ 1-\tanh\left(\frac{R-R_{\mrm{DZ}}}{\Delta_{\mrm{DZ}}}\right)\right],
    \label{eq:dead zone}
\end{equation}
where $\alpha_0 = 10^{-3}$ and $\alpha_{\mrm{DZ}} = 10^{-5}$ are the Shakura \& Sunyaev viscosity parameters outside and inside the dead zone. The viscosity transition is located at $R_\mrm{DZ} = 1.5R_0$ and the width of the transition region is $\Delta_{\mrm{DZ}}$.

Due to the low viscosity within the dead zone, gas gradually piles at the dead zone edge to form a density bump. If $\Delta_\mrm{DZ}<2H$, the RWI will be triggered to generate large-scale anticyclonic vortices~\citep{regaly2011possible}.

In this paper, we set up five models to investigate how the viscosity transition region width (SD {\it vs} H-SH {\it vs} Q-SH) and the disk self-gravity (SD {\it vs} HM-$10\mrm{M_J}$ {\it vs} HM-$30\mrm{M_J}$) affect vortex-driven spiral arms. The setup of the models is in Table~\ref{tab:model}. We run 3000 orbits for each model in our hydrodynamic simulations.

\begin{table}[!htbp]
\centering
\caption{List of Models}
\label{tab:model}
\begin{tabular}{cccc}
\toprule
Name  & $\mrm{\Delta_\mrm{DZ}}$  & Self-Gravity & Disk Mass $(\mrm{M_J})$ \\
\midrule
SD               & $H  $ & No &   3  \\
H-SH              & $H/2$ & No &   3  \\
Q-SH              & $H/4$ & No &   3  \\
HM-$10\mrm{M_J}$ & $H  $ & Yes & 10 \\
HM-$30\mrm{M_J}$ & $H  $ & Yes & 30 \\
\bottomrule
\end{tabular}
\\
SD: Standard\\
H-SH: Half-Scale Height\\
Q-SH: Quarter-Scale Height\\
HM-$10\mrm{M_J}$: High Disk Mass with $M_{\rm disk}=10\;\mrm{M_J}$\\
HM-$30\mrm{M_J}$: High Disk Mass with $M_{\rm disk}=30\;\mrm{M_J}$\\
\end{table}

\section{The Vortices in the Models}\label{sec:hydro_results}
\subsection{The Standard Model (SD)}
\begin{figure*}[t]
\centering
\includegraphics[width=1.0\textwidth]{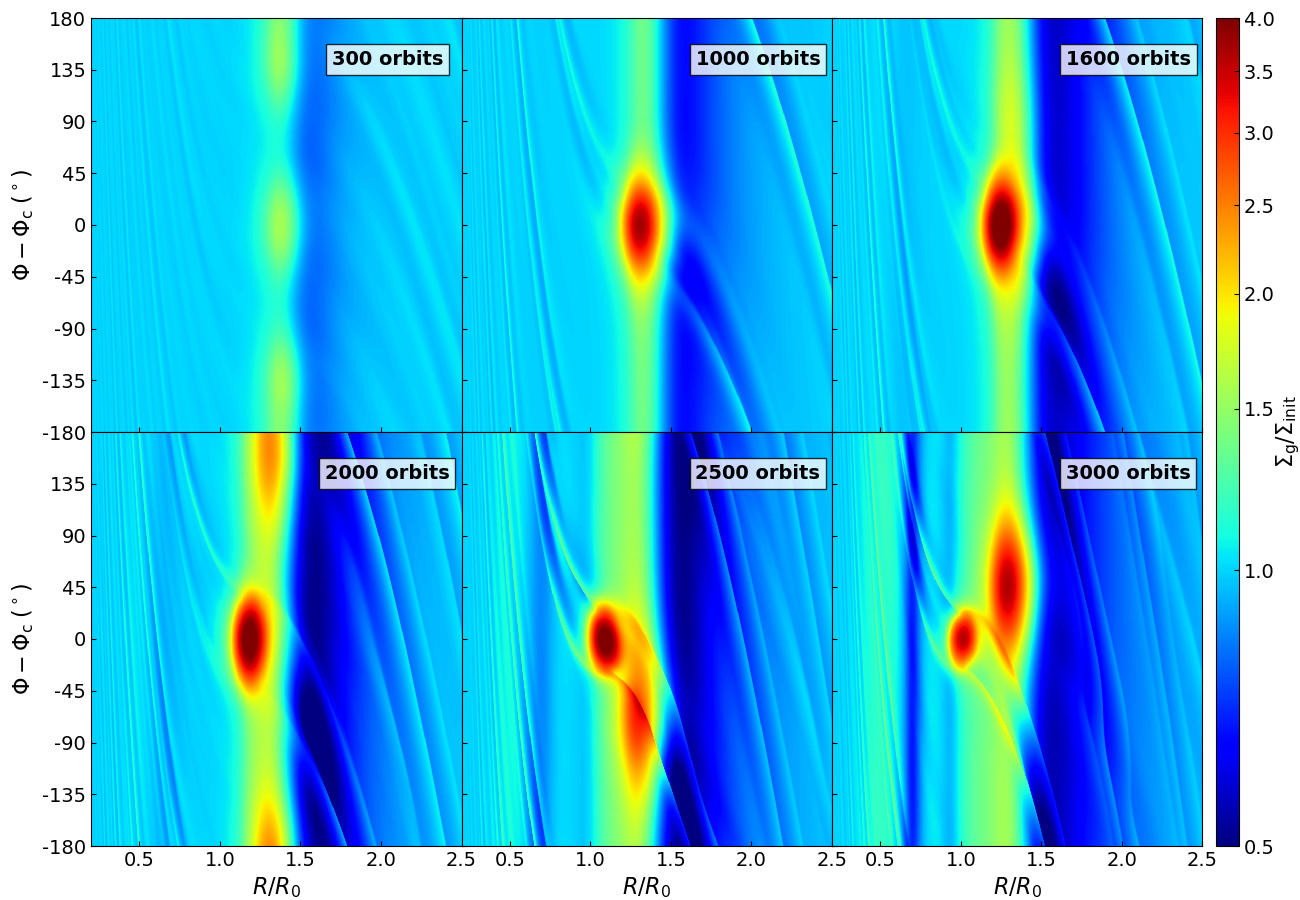}
\caption{The surface density profile normalized by $\Sigma_\text{init}$ at various epochs for Model SD. $\Sigma_\text{init}$ is the initial surface density for each model, i.e.\ the surface density for each model at 0 orbit. The horizontal axis is the radius, and the vertical axis is the azimuth (the center of the first generation vortex is at $\Phi_\mrm{c}=0$). We define the vortex center where the velocity in the vortex equals to the local Keplerian velocity. The centers of vortices are very close to the their density maxima.}
\label{fig:Density_SD}
\end{figure*}

\begin{figure*}[t]
\centering
\includegraphics[width=1.0\textwidth]{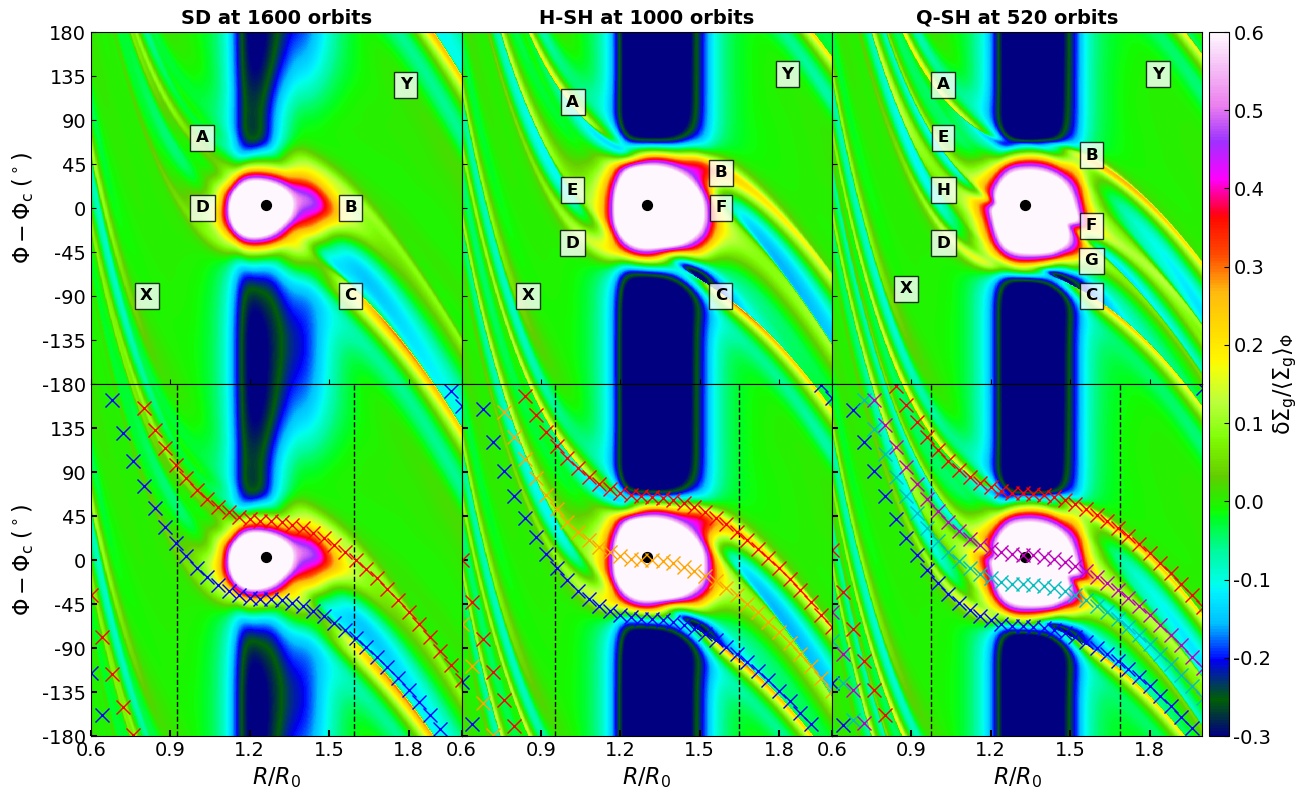}
\caption{The surface density contrast defined by Equation~\ref{eq:contrast} of the models with various viscosity transition widths (SD, H-SH and Q-SH). The two rows are the same expect for the mark ups and symbols. The black solid points mark the centers of the vortices. In the top row, the spirals are marked using letters (primary spirals A-H; secondary spirals X-Y). In the bottom row, the cross symbols mark the expected wake of the primary spirals in the linear density wave theory described by Equation~\ref{eq:shape} , and the vertical dashed lines mark the radial locations 5 scale heights away from the vortex centers. The azimuthal cuts at these locations will be shown in Figure~\ref{fig:wakes}.}
\label{fig:Density_viscosity}
\end{figure*}

Figure~\ref{fig:Density_SD} shows the evolution of the gas surface density in Model SD at various epochs. Due to the low viscosity in the dead zone region, the gas piles up at the dead zone edge ($\sim 1.3R_0$). Prior to the violation of the Rayleigh criterion ($\kappa^2\equiv \frac{1}{R^3} \frac{d}{dR}\left(R^4 \Omega^2\right) < 0$, where $\kappa$ is the epicyclic frequency), the RWI condition~\citep{lovelace1999rossby} is violated first. The RWI is then triggered, and generates large scale vortices~\citep{li2000rossby,li2001rossby,ono2016parametric,ono2018parametric}. The density bump in Model SD becomes Rossby wave unstable at 300 orbits (upper left panel), then breaks into three mode-3 vortices. They then merge into a single vortex at about 600 orbits. There are two pairs of spirals connected with the edges of the vortex (e.g., at 1000 orbits). The spirals protect the vortex from being smeared out by the background Keplerian flow~\citep{li2001rossby}. Before a vortex is smeared out, the vortex grows stronger by accreting gas (1600 orbits).

A vortex is part of the disk, and it generates spiral density waves by perturbing the velocity in the background flow~\citep{bodo2005spiral,heinemann2009excitation}. Vortices generate density waves by perturbing the background flow and producing sound waves, which are sheared by the Keplerian rotation. There is no planet and no disk self-gravity in Model SD, so the wave excitation mechanism is unrelated to gravitational interaction between vortices and the disk --- the vortices are massless. This is different from how planets excite density waves, in which case a planet is a point mass and launches density waves at Lindblad resonances through gravitational disk-planet interactions~\citep{goldreich1979excitation}.

\cite{van2016vortices} showed the ALMA Band 7 continuum observations on SAO~206462. They found that a previously observed asymmetric ring in SAO~206462 is consisted of an inner ring and an outer bump, and proposed the possibility that the latter is a vortex driving the observed spiral arms in scattered light~\citep{muto2012discovery}.
\cite{cazzoletti2018evidence} presented multi-wavelengths ALMA observations of the dust continuum from the SAO~206462 circumstellar disk. They resolved the disk into an axisymmetric ring and a crescent. They claimed that the crescent is a massive vortex with several $\mrm{M_J}$ and it launches the observed spirals arms in a way similar to the interactions between a disk and a massive planet. This interpretation is inconsistent with spirals driven by vortices. In addition to the origin, the contrast of spirals induced by vortices is too low to be detectable in scattered light. We will explain it in Section~\ref{sec:strength}.

The density waves exert torques onto the vortex. At 2000 orbits, the negative torque by the outer density waves surpasses the positive torque by the inner density waves due to the radial asymmetry in the disk~\citep{paardekooper2010vortex}. As a result, the vortex migrates inward. The vortex in Model SD forms at about $1.3R_0$, launching spiral waves and migrating to $\sim1.0R_0$ at the end of our simulation (3000 orbits). The migration of a vortex is similar to the type I migration of a planet~\citep{paardekooper2010vortex}. After that, gas keep piling up at the dead zone edge, triggering the RWI to generate a secondary generation of vortices (the outer vortex seen at 2000 orbits and later). However, we do not see significant spirals excited by the second generation of vortices due to their small velocity perturbation on the background and a larger aspect ratio~\citep{surville2015quasi}.

At 3000 orbits, there are two shallow gaps at $R\sim0.7$ and 0.9 opened by the spirals driven by the first generation vortex in the dead zone. \cite{dong2017multiple,dong2018multiple} and \citet{bae2017formation} showed that a Super Earth opens multiple gaps in low viscosity disks with $\alpha \lesssim 10^{-4}$. It is possible that the vortex in Model SD opens multiple gaps in a similar way.

Based on \cite{meheut2013strong}, the RWI is triggered by the local Rossby waves trapped in the disk. The Rossby wave is a potential vorticity (a.k.a. vortensity) wave. The vortensity is conserved along the streamlines in an inviscid and barotropic fluid. In the RWI, the enthalpy perturbation $\psi \equiv \delta P/\Sigma$ is similar with the Schr\"odinger equation~\citep{lovelace2014rossby}. When Rossby waves escape the potential well, they transfer into spiral density waves propagating inward and outward. This phenomenon is similar with the ``tunnel through'' in quantum mechanics~\citep{li2001rossby}. The spiral density waves cause the dissipation and shrinking of the vortex. The primary vortex at 3000 orbits in Figure~\ref{fig:Density_SD} is weaker than that at 2500 orbits due to this effect.

\subsection{The Effects of Viscosity Transition Width}
Figure~\ref{fig:Density_viscosity} show the vortices and spirals in Model SD, H-SH and Q-SH. We pick the orbital frames prior to the production of the secondary generation vortices. By comparing models with different radial widths of the viscosity transition region, we find that the Rossby wave unstable density bump is replenished faster when the region shrinks. The growth rate of the RWI becomes higher while the lifetime of the resulting vortices is shorter. The primary vortex in Model SD survives for at least 3000 orbits. The vortices in Model H-SH and Q-SH only survive for 2000 and 1000 orbits, respectively. After that, the vortices are smeared out by the background Keplerian flow into axisymmetric density bumps. Different from Model SD, there is no secondary vortex in H-SH and Q-SH. Based on \cite{surville2013structure}, more elongated vortices migrate more slowly, and smaller vortices are more compressible. The asymmetries between the inner and outer spirals in smaller vortices lead faster migrated rate.

\subsection{The Effects of Self-Gravity and Disk Mass}

\cite{lin2011effect} found that disk self-gravity can stabilize high mode number vortices. This is consistent with our high mass self-gravitating models --- at 1600 orbits, there are two mode-2 vortices in Model HM-$10\mrm{M_J}$, four mode-4 vortices in Model HM-$30\mrm{M_J}$, but only one vortex (mode-1) in Model SD (Figure~\ref{fig:Density_gravity}). These vortices in HM-$10\mrm{M_J}$ and HM-$30\mrm{M_J}$ at 1600 orbits survive to the end of the simulations (3000 orbits). Comparing Model SD, HM-$10\mrm{M_J}$ and HM-$30\mrm{M_J}$, the effect of disk self-gravity on the vortices becomes larger when the disk becomes more massive. The Toomre Q~$\equiv \frac{\kappa c_\mrm{s}}{\pi G \Sigma}$ \citep{toomre1964gravitational} reaches its maximum at the location of the vortex in all models. Q is about 20 and 10 in Models HM-$10\mrm{M_J}$ and HM-$30\mrm{M_J}$ at 1600 orbits, respectively, while $Q>50$ in the other three viscosity transition models by the end of the simulations (3000 orbits). Based on \cite{lovelace2012rossby}, disk self-gravity is important for the RWI only when $Q<R/H$. In our models, the disk self-gravity becomes significant when Q $<R/H\sim$20 at the locations of the vortices.

\section{Properties of Spirals}\label{sec:properties}

\subsection{Number of Spirals}
\begin{figure*}[t]
\centering
\includegraphics[width=1.0\textwidth]{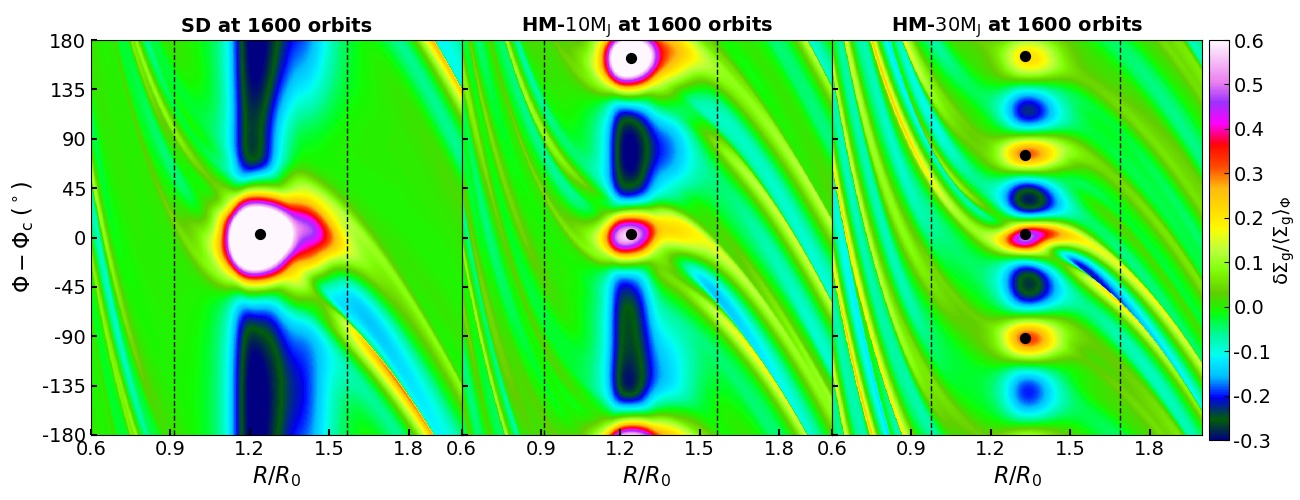}
\caption{The surface density contrast defined by Equation~\ref{eq:contrast} of the models with self-gravity (HM-$10\mrm{M_J}$ and HM-$30\mrm{M_J}$) comparing with Model SD (no self-gravity). The black solid points mark the centers of the vortices. The vertical dashed lines mark the radial locations 5 scale heights away from the vortex centers; the azimuthal cuts at these locations will be shown in Figure~\ref{fig:wakes}.}
\label{fig:Density_gravity}
\end{figure*}

Figures~\ref{fig:Density_viscosity} and \ref{fig:Density_gravity} show the general morphology and contrast of the vortices and spirals at various epochs for different models. There are 4, 6, and 8 spirals connected with the vortex in Model SD (A-D), H-SH (A-F), and Q-SH (A-H), respectively. \cite{li2001rossby} showed that in the case of four spirals, two of them (A and C) are shocks, while the other two (B and D) are rarefaction waves. The vortices of Model H-SH and Q-SH are more elongated than the vortex in Model SD. When a vortex is too long (aspect ratio $\gtrsim 9$ in our simulations), it can not communicate along the $\Phi$ direction. An elongated vortex will either lead to multiple density peaks or production of multiple spirals. Multiple density peaks can be the outcome of incomplete vortex merger in the initial formation stage. The spirals E and F in H-SH are produced from spirals D and B. The vortex in Q-SH has two density peaks (i.e. two cores), so that it excites eight spirals. In Model HM-$10\mrm{M_J}$ and HM-$30\mrm{M_J}$, spiral arms induced by different vortices start to interfere with each other (Figure~\ref{fig:Density_gravity}), making identifying each unique arm difficult.

In the past, a sub-thermal mass planet was thought to generate a pair of spiral arms (one on each side of the planet's orbit; \citealt{ogilvie2002wake}). However, based on the simulations of \cite{bae2018planet1} and \cite{miranda2019multiple} (see also \citealt{zhu2015structure,bae2017formation,bae2018planet2,fung2015inferring,lee2016ultraharmonics}), the angular momentum flux of the primary density waves induced by a sub-thermal mass planet can be transferred into higher-order density waves (secondary, tertiary, etc) during their propagation. In our simulations, the secondary spirals (spirals X and Y) are not originated at the vortices, and they are likely the higher order spirals generated by the primary spirals (spirals A and C) as a result of the constructive interference among different azimuthal modes~\citep{bae2018planet1,miranda2019multiple}.

\subsection{Shape of Spirals}
Based on the linear spiral density wave theory~\citep{goldreich1979excitation,rafikov2002planet,muto2012discovery}, the shape of spirals induced by a planet in the linear regime is:
\begin{equation}
    \begin{array}{lll}
    {\Phi\left(R\right) = \Phi_\mrm{e} - \frac{\mrm{sgn}\left(R - R_\mrm{e}\right)}{H_\mrm{e}}}\\
    {\times\left\{\left(\frac{R}{R_\mrm{e}}\right)^{1+\beta}\left[\frac{1}{1+\beta} - \frac{1}{1-\alpha +\beta}\left(\frac{R}{R_\mrm{e}}\right)^{-\alpha}\right]
    -\left(\frac{1}{1+\beta} - \frac{1}{1-\alpha+\beta}\right)\right\}}
    \end{array}
    \label{eq:shape}
\end{equation}
where the orbital angular frequency $\Omega \propto R^{-\alpha}$ and the sound speed $c_\mrm{s} \propto R^{-\beta}$ ($\alpha = \frac{3}{2}$ and $\beta = \frac{1}{4}$ in our simulations). $R_\mrm{e}$ and $\Phi_\mrm{e}$ are the radial and azimuthal locations of the planet. When the mass of a planet is larger than the thermal mass, the linear theory breaks down and there is no analytic theory for the shape of the spirals.

Equation~\ref{eq:shape} is marked using cross symbols in Figure~\ref{fig:Density_viscosity} for Model SD, H-SH, and Q-SH. The shapes of the spirals A-D in these models are consistent with Equation~\ref{eq:shape}. However, as the multiple spiral arms are likely interacting with each others, there are deviations in the shapes of the spirals E-H. As mentioned before, spirals induced by a vortex are similar to those induced by a sub-thermal mass planet. After the excitation of the density waves, the specific excitation mechanism no longer affects the waves as they propagate. The spirals X and Y are not connected with the vortices; it is harder to compare their shapes with the linear theory due to the uncertainty in the origin of the spirals.

\subsection{Contrast of Spirals}\label{sec:strength}
\begin{figure*}[t]
\centering
\includegraphics[width=0.95\textwidth]{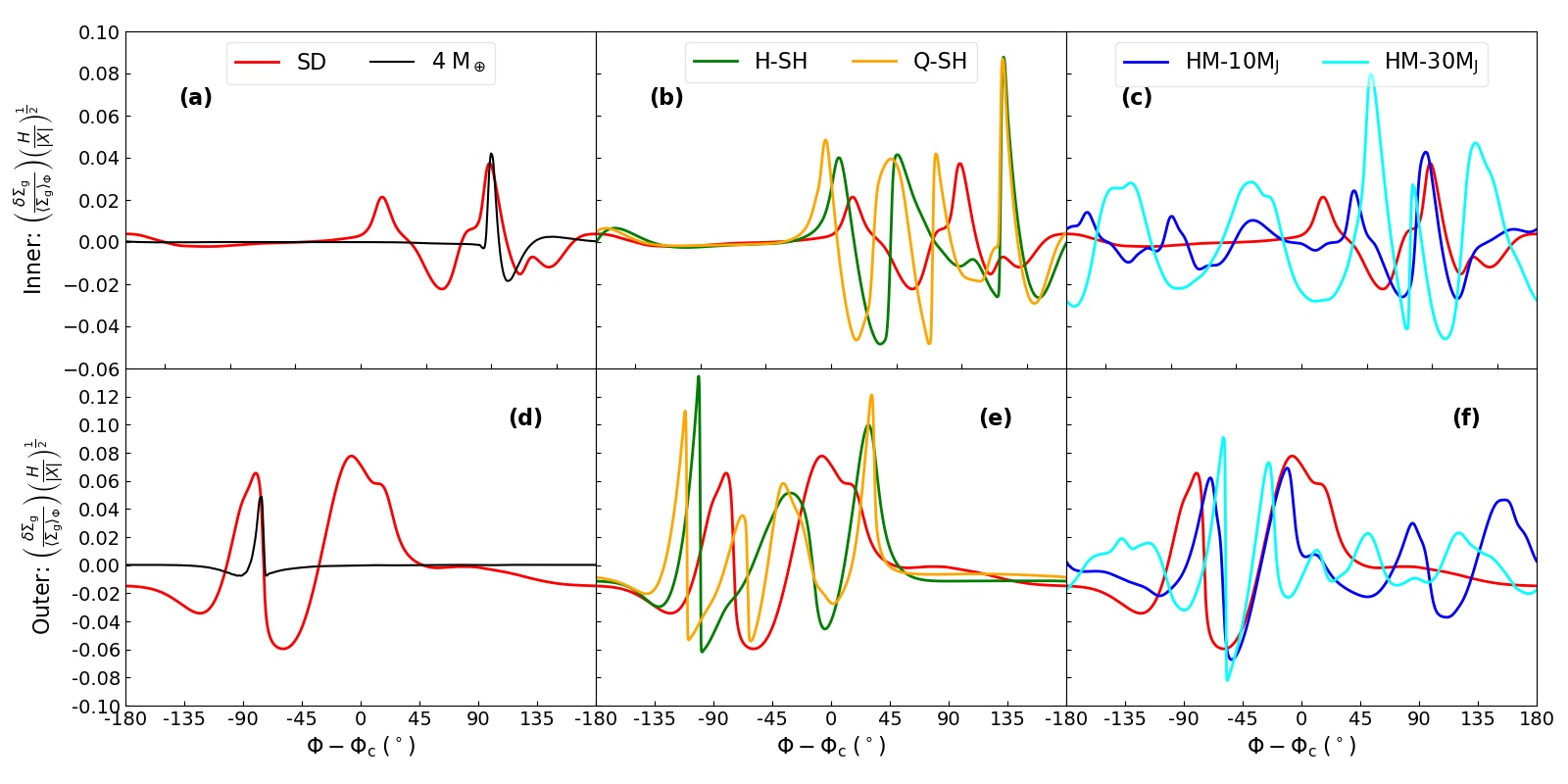}
\caption{Azimuthal density profiles across the spirals at a radial distance 5 scale heights away from the vortex center (the vertical dashed lines in Figures~\ref{fig:Density_viscosity} and~\ref{fig:Density_gravity}). The horizontal axis is the azimuth. The vertical axis is the surface density contrast normalized by $\left(|X|/H\right)^{1/2}$, where $X \equiv R - R_\mrm{c}$, and $R_\mrm{c}$ is the radial location of the vortex center. The upper and lower rows are for the inner (at $R=R_\mrm{c}-5H$) and outer spirals (at $R=R_\mrm{c}+5H$), respectively. The black lines in panels ($a$) and ($d$) are the azimuthal cuts of the spiral waves generated by a $4\;\mrm{M_\oplus}$ planet in the $H/R = 0.05$ planet-disk interaction simulation in~\cite{dong2017multiple}.}
\label{fig:wakes}
\end{figure*}

The contrast of density waves in gas surface density at a given radius $R$ is defined as:
\begin{equation}
\frac{\delta\Sigma_\text{g}\left(R\right)}{\langle\Sigma_\text{g}\left(R\right)\rangle_\Phi}\equiv\frac{\Sigma_\text{g,peak}}{\langle \Sigma_\text{g}\left(R\right)\rangle_\Phi}-1
    \label{eq:contrast}
\end{equation}
where $\langle \Sigma_\text{g}\left(R\right)\rangle_\Phi$ is the $2\pi$ azimuthally averaged surface density at $R$. Figure~\ref{fig:wakes} shows the azimuthal cuts of the scaled surface density at 5 scale heights from the vortex centers, and highlights how the contrast depends on the width of the viscosity transition (panels $b$ and $e$) and self-gravity and disk mass (panels $c$ and $f$). We note that the outer spirals are stronger than the inner ones.

To compare the spirals induced by vortices and planets, we show the same azimuthal cuts of the spiral waves driven by a $4\;\mrm{M_\oplus}$ planet in an $H/R=0.05$ and $\alpha = 5 \times 10^{-5}$ disk~\citep{dong2017multiple} in panels ($a$) and ($d$). The difference in the contrast between the inner and outer spirals is smaller in the case of planet-induced spirals than vortex-induced spirals. This leads to higher migration rates of vortices than planets~\citep{paardekooper2010vortex}. In planet-disk interaction, we can define the thermal mass as $M_\text{th} = \left(H/R\right)^3 M_\star$~\citep{goodman2001planetary}, which is about $50\;\mrm{M_\oplus}$ at $1.3R_0$ in our simulations. For planet-induced spiral arms, the peak of the scaled surface density contrast is $\left(\frac{\delta\Sigma_\mrm{g}}{\langle \Sigma_\mrm{g}\rangle_\Phi}\right)\left(\frac{H}{|X|}\right)^\frac{1}{2}\approx M_\text{planet}/M_\text{th}$ in the linear regime ($M_\text{planet}\ll M_\text{th}$), where $X \equiv R - R_\mrm{c}$ and $R_\mrm{c}$ is the radius of the vortex peak~\citep{goodman2001planetary,dong2011density1,duffell2012global}. The surface density contrasts of the spirals produced by the vortices in our simulations are similar to those induced by a sub-thermal mass planet. As an example, the contrast of the spirals wakes in Model SD are similar with that of the spirals induced by a $4\;\mrm{M_\oplus}$ planet ($\sim0.1\;M_\text{th}$).

The density bumps in H-SH and Q-SH are more Rossby wave unstable than that in SD due to the steeper viscosity transition compared with SD. The vortensity of vortices in H-SH and Q-SH are about 1.5 and 2.0 times the vortensity of the primary vortex in SD. Because the vortensity represents the rotation of vortices with respect to the background flow, the velocity perturbations and contrast of the vortices in H-SH and Q-SH are larger and stronger than those in SD.

Model HM-$10\mrm{M_J}$ and HM-$30\mrm{M_J}$ each produces multiple long-lasting vortices. Specifically, the vortex located at $0^\circ$ has a shorter rotation period and a lower aspect ratio, the contrasts of its spirals are higher than those induced by the other more extended vortices~\citep{bodo2005spiral,surville2015quasi}. While the disk self-gravity compress the gas, the contrast of vortex-induced spirals is insensitive to the disk self-gravity and the mass of the vortex. The vortex in Model HM-$10\mrm{M_J}$ and Model HM-$30\mrm{M_J}$ weights $\sim 0.8\; \mrm{M_J}$ and $\sim 1.2\; \mrm{M_J}$, respectively (5.0 and 7.5 times the thermal mass); however, the peak contrasts of their density waves are within a factor of 2 from those in Model SD, which has a massless vortex (see also \S\ref{sec:strength}).

The contrast of the spirals in our models is in between 0.1 and 0.3. \cite{juhasz2015spiral} and \cite{dong2017bright} showed that when the surface density contrast of spiral arms is smaller than order unity (corresponding to spirals driven by sub-$ \mth$ planets), it is difficult to detect the spirals at tens of AU in systems at 140 pc under the angular resolution currently achievable in NIR scattered light imaging.

\section{Conclusions}\label{sec:summary}
We have run hydrodynamic simulations to investigate the properties of spiral arms induced by a vortex. Here are the main conclusions of this paper:
\begin{enumerate}
    \item A massless vortex can generate spiral arms as its velocity field compresses the background gas to produce density waves. This is different from how planets generate spiral arms through gravitational planet-disk interactions.
    \item The surface density contrast of vortex-driven spirals is from $0.1$ to $0.3$ in our simulations. This is similar to the spirals produced by a planet with a mass on the order of 0.1 thermal mass, equivalent to a few to a few tens of Earth masses at tens of AU under typical conditions. Such arms are too weak and should not be expected to be detectable in current direct imaging observations. The prominent spirals observed in protoplanetary disks such as MWC~758, SAO~206462, LkH$\alpha$~330 are unlikely to be driven by the candidate vortex seen in them.
    \item The disk self-gravity becomes important to the development of the Rossby Wave Instability when $Q\lesssim R/H$, in which case it stabilizes the high mode number vortices produced by the RWI. The surface density contrast of vortex-driven spirals is insensitive to the disk self-gravity. Specifically, the contrasts of the spiral arms driven by the vortices in Model HM-30$\rm M_J$, which weight about $1.2\; \mrm{M_J}$ (7.5 times the disk thermal mass), are still comparable to those driven by sub-thermal mass planets.
    \item A vortex generates at least 4 spiral arms (two on each side) that are directly connected with itself, and the shape of these spirals are consistent with the linear density wave theory. More elongated vortex can generate more than 4 spirals. A vortex can produce secondary spirals in the disk.
\end{enumerate}


\section*{Acknowledgments}
We thank the referee for the detailed comments that improved the presentation of this paper significantly. We are grateful to Jaehan Bae, Tomohiro Ono, Min-Kai Lin and Chong Yu for useful discussions. This work is supported by the National Natural Science Foundation of China (grant Nos. 11773081, 11661161013, 11633009 and 11873097), the CAS Interdisciplinary Innovation Team, the Strategic Priority Research Program on Space Science, the Chinese Academy of Sciences, Grant No. XDA15020302 and the Foundation of Minor Planets of Purple Mountain Observatory. We also acknowledge the support by a LANL/CSES project. This work was partially performed at the Aspen Center for Physics, which is supported by National Science Foundation grant PHY-1607611. 

\bibliography{references}
\end{document}